# TOWARD A VIRTUAL MATERIAL FOR LIFETIME PREDICTION OF CMCs

M. Genet[1], P. Ladevèze[12], G. Lubineau[1], E. Baranger[1], A. Mouret[3]

[1] LMT-Cachan (ENS-Cachan, CNRS, Paris 6 University, UniverSud Paris PRES), 61 Avenue du Président Wilson, 94235 Cachan Cedex, France,
{genet, ladeveze, lubineau, baranger}@lmt.ens-cachan.fr
[2] EADS Foundation Chair, Advanced Computational Structural Mechanics
[3] Snecma Propulsion Solide, SAFRAN Group, Les 5 chemins, 33187 Le Haillan Cedex, France,
anne.mouret@snecma.fr

## ABSTRACT

A first version of a multi-scale, multi-physic & hybrid model –called virtual material– for predictions on Self-Healing Ceramic Matrix Composite's (CMCs) lifetime is presented. The model has a mechanical and a chemical part, which are presented here in their actual state of development. The mechanical part provides precise data for the chemical models through an hybrid –melting continuum damage macro-model & discrete crack surfaces– representation of the morphology of the crack network at yarn scale. The chemical part should provide predictions on the structure's lifetime using a model of the self-healing process, not yet achieved then not presented here, and a model of fiber sub-critical failure under mechanical and chemical load.

## 1. INTRODUCTION

Snecma Propulsion Solide, SAFRAN group company, has developed a range of woven SiC/[Si-B-C] composites designated as self-healing materials [1]. The composites are build up from woven yarns of SiC fibers infiltrated by a multi-layered ceramic matrix. During the load, a first crack network, perpendicular to the principal traction, appears in the matrix between the yarns [2] (see Fig. 1). Once it is saturated, a second network, oriented by the fibers, appears in the matrix within the yarns without breaking the fibers thanks to a PyC interphase between fibers and matrix [2, 3] (see Fig. 1).

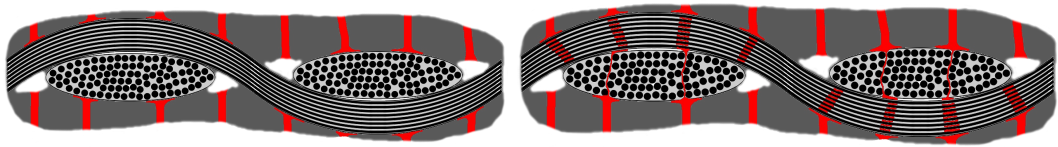

Figure 1: Crack network.

The self-healing process consists in filling these cracks with an oxide resulting from oxidation of certain components of the matrix, which limits the diffusion of oxygen toward fibers, that are the core of the material and might suffer from sub-critical cracking under oxidizing atmosphere [4]. Therefore, this process leads to a great increase in the material's lifetime. The challenge associated with the use of such materials for large industrial applications resides in the development of robust models of their mechanical and chemical behaviour up to failure, based on the physics of their micro-mechanisms so as to extrapolate their response even for very large lifetimes.

A first macro-model has been developed at Cachan [5] that predict the mechanical behavior of the composites with an a priori damage kinematics. This macro-model has been improved with non imposed damage kinematics [6] and then with the introduction of chemical mechanisms [7]. It is still under development [8]. However, as a macro-model, this model is limited in the description of the micro-mechanisms, although they play a major role in the material's lifetime.

The goal of this paper is to present a first version of our proposed multi-scale, multi-physic & hybrid model -called virtual material- for lifetime predictions on CMCs. This model has two main parts, which are presented here in their actual state of development. The mechanical part is presented section 2. It is based on the geometry of the material at yarn scale, and provides a representation of the morphology of the crack network in an hybrid way so as to focus on essential informations:

- Inter-yarn matrix cracks, having no quantitative effect on the chemical micro-mechanisms, are modelled using a continuum anisotropic damage macro-model [5],
- Intra-yarn matrix cracks, having a substantial effect on both mechanical – through the debonding zone induced between fibers and matrix [ACK]– and chemical –through the influence on the amount of oxygen arriving around the fibers [8]– micro-mechanisms, are modelled using discrete crack surfaces [9, 10], whose behaviour is obtained though a micro-macro link with a micro-model of the fiber/matrix debonding [11, 7].

The chemical part is presented section 3. It is based on a model of the stress and oxidation induced sub-critical growth of the surface defects of the fibers toward their core until fracture [12, 4], which is presented here. The self-healing mechanisms [7, 8] will be handled later on. This part should then provide lifetime predictions on the structure through time to rupture predictions on fibers, boundary conditions of sub-critical crack growth and self-healing models (applied stress, opening of the cracks, etc) being provided by the mechanical part.

## 2. MECHANICAL PART OF THE MODEL

The mechanical behaviour of the material is studied using finite-element method on a meso-cell at yarn scale, which is constituted as follow:

- The yarns, which behaviour is obtained by homogenization of a micro-cell at fibre scale,
- The intra-yarns cracks, that appear directly in the calculation through the use of potential fracture surfaces,
- The eventually damaged matrix between yarns, where cracks are taken into account through the use of an anisotropic damage behaviour model.

The way the internal geometry of the woven yarns is taken into account in the model is presented section 2.1. The potential fracture surfaces and the associated formulation are presented section 2.2. The damage macro-model is presented section 2.3.

### 2.1 INTERNAL GEOMETRY

Three points have to be taken into account when dealing with the geometry of woven composites at yarn scale (see Fig. 2):

- The geometry of the woven yarns is complex in itself,
- The geometry of the matrix volume between yarns, including porosity, is even more complex,
- All the geometrical parameters (woven sequence, matrix thickness, etc) are not known exactly and contain a lot of variability.

Some people in the literature are developing tools to represent exactly this geometry [13, 14], and are facing associated problems: meshing complex volumes, identifying many geometrical parameters, etc.

The choice made here leads to a debased representation of the geometry but avoids many technical problems: regular meshes –then not conform to the geometry– are used, on which are defined level set functions so as to determine to which entity –yarns, matrix or voids– belongs each geometrical point, and an integration method [leclerc] computes the associated stiffness matrix. Taken into account the variability in the geometry that has to be meshed, this method will be assumed fine enough for the presented model.

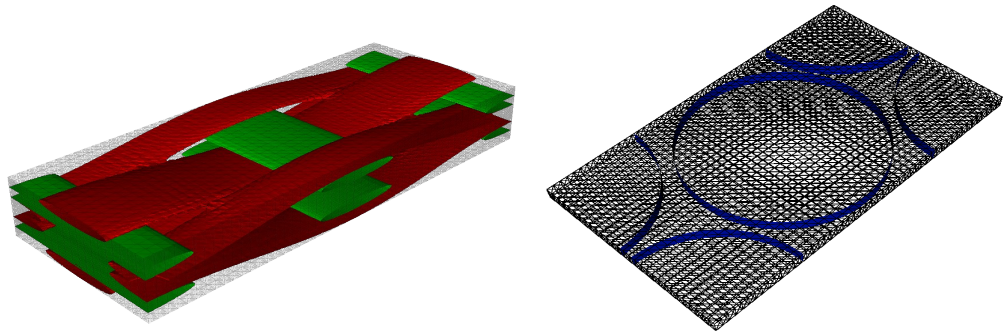

Figure 2: Representation of the geometry at yarn and fiber scale.

## 2.2 POTENTIAL CRACK SURFACES

Three points have to be taken into account when dealing with the intra-longitudinal yarns matrix cracks:
- Their direction is known a priori: they are orthogonal to fibers,
- They do not break the fibers, because of the PyC interphase [3], so there is still a stiffness in the material at the location of a crack,
- The behavior of the crack (opening, inelastic deformation, etc) is driven by the fiber/matrix debonding process [3, 11].

As the topology of this crack network is known a priori, it is possible to mesh the cracks when creating the meso-cell. Some 2D examples –3D has not yet been coded– created using GMSH [15] are given Fig. 3.

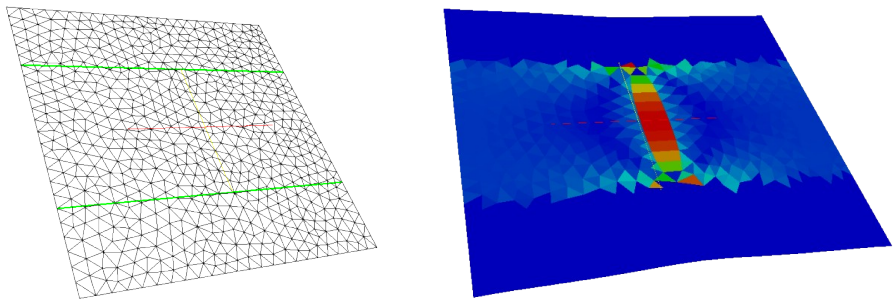

Figure 3: 2D Crack surfaces, meshes and associated stress field.

The formulation associated with opened crack surfaces is given by the micro-macro link described now. Let us consider the following two problems:

- A fiber, surrounded by a matrix, a transverse crack in the matrix, a shear-lag type load sharing [7, 11] –the shear stress is written $\tau$, the debonding zone size is written $l_d$–, problem 1 (see Fig. 4a),
- Two fiber-matrix material parts separated by a cohesive zone, problem 2 (see Fig. 4b).

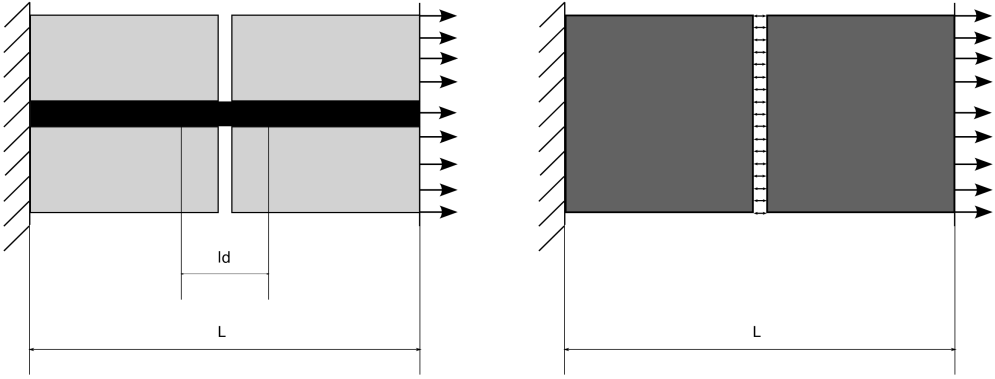

Figure 4: Two equivalent problems, shear lag model and cohesive zone model.

The energy of the shear lag model is:
$$\begin{aligned}E_d^{shear\ lag} &= E_d^f + E_d^m \\ &= 2S_f \int_0^{L/2-l_d} \frac{1}{2}\frac{\sigma_f^2}{E_f} + 2S_f \int_{L/2-l_d}^{L/2} \frac{1}{2}\frac{\sigma_f^2}{E_f} \\ &+ 2S_m \int_0^{L/2-l_d} \frac{1}{2}\frac{\sigma_m^2}{E_m} + 2S_m \int_{L/2-l_d}^{L/2} \frac{1}{2}\frac{\sigma_m^2}{E_m} \\ &= \frac{\sigma_f^2 S_f (L/2 - l_d)}{E_f} + \frac{S_f}{E_f}\left(\frac{\sigma^2 l_d}{V_f^2} - \frac{\sigma \tau l_d^2}{V_f r_f} + \frac{\tau^2 l_d^3}{r_f^2}\right) \\ &+ \frac{\sigma_m^2 S_m (L/2 - l_d)}{E_m} + \frac{\sigma^2 E_m S_m l_d}{2 E_{moy}^2}\end{aligned}$$

Where $S$ is the considered section, $\sigma$ the applied stress, $E_{moy} = E_f V_f + E_m V_m$ the Young modulus of the fiber-matrix material, $V_f$ & $V_m$ being the portion of fiber and matrix in the cell.

The energy of the cohesive zone model is:
$$\begin{aligned}E_d^{cohesion} &= 2S \int_0^L \frac{1}{2}\frac{\sigma^2}{E_{moy}} + \frac{1}{2}\frac{\sigma^2 S^2}{k} \\ &= \frac{\sigma^2 S L}{E_{moy}} + \frac{\sigma^2 S^2}{2k}\end{aligned}$$

Where $k$ is the stiffness of the cohesive zone.

Equating the two energies, the cohesive zone stiffness is obtained as a function of the micro physical parameters, is independent of the size $L$ of the above problems, but depends on the applied stress $\sigma$, which means that the behavior of the cohesive zone is non linear, even in traction.

## 2.3 ANISOTROPIC DAMAGE MACRO-MODEL
The important points concerning the inter-yarn matrix cracking are the following:

- The direction of the crack network is not known a priori: it is oriented by the external load,
- When the material is in compression, the cracks are closed and the behavior is healthy if the cracks are in pure compression and damaged if the cracks support shear.

The macro-model used has been build in the framework of the anisotropic damage theory [6], and is already used as a complete macro-model –with inter- and intra-yarn matrix cracks– [5, 6, 7, 8]. It is used here for the inter-yarn matrix cracks only. The main idea is to split the elastic energy into a 'traction' energy, a 'compression' energy and an energy identical for traction and compression:

$$e_d = e_t + e_c + e_{tc} \text{, with } \begin{cases} e_t = \frac{1}{2}\left(\underline{\underline{\sigma^+}} : \underline{\underline{\underline{\underline{C}}}} : \underline{\underline{\sigma^+}}\right) \\ e_c = \frac{1}{2}\left(\underline{\underline{\sigma_-}} : \underline{\underline{\underline{\underline{C_0}}}} : \underline{\underline{\sigma_-}}\right) \\ e_{tc} = \frac{1}{2}\left(\underline{\underline{\sigma}} : \underline{\underline{\underline{\underline{Z}}}} : \underline{\underline{\sigma}}\right) \end{cases}$$

Where $\underline{\underline{\underline{\underline{C_0}}}}$ is the initial compliance operator, $\underline{\underline{\underline{\underline{C}}}}$ is the damaged compliance operator for traction –it is an internal variable of the model, initially equal to $\underline{\underline{\underline{\underline{C_0}}}}$–, $\underline{\underline{\underline{\underline{Z}}}}$ the compliance operator of the damage part independent of the traction/compression state –another internal variable of the model, initially null–. Moreover, $\underline{\underline{\sigma^+}}$ is the positive part of $\underline{\underline{\sigma}}$ affected by the damage state, and $\underline{\underline{\sigma_-}}$ the negative part of $\underline{\underline{\sigma}}$ in the classical sense – because there is no damage in compression–:

$$\begin{cases} \underline{\underline{\sigma^+}} = \underline{\underline{\underline{\underline{H}}}}^{-1} \langle \underline{\underline{\underline{\underline{H}}}} \, \underline{\underline{\sigma}} \rangle_+ \\ \underline{\underline{\sigma_-}} = \langle \underline{\underline{\sigma}} \rangle_- \end{cases} \text{, with } \underline{\underline{\underline{\underline{H}}}}^2 = \underline{\underline{\underline{\underline{C}}}}$$

The differentiability of such an energy has been proved in [lad 84]. The damage kinematics is totally defined by the damage evolutions laws, which are build from the following damage forces:

$$\begin{cases} \underline{\underline{\underline{\underline{Y_C}}}} = \frac{\partial e_d}{\partial \underline{\underline{\underline{\underline{C}}}}} = \underline{\underline{\sigma^+}} \otimes \underline{\underline{\sigma^+}} \\ \underline{\underline{\underline{\underline{Y''}}}} = \underline{\underline{L}} \otimes \underline{\underline{L}} \end{cases} \text{, with } \underline{\underline{L}} = \left(\underline{\underline{\underline{\underline{i_2}}}} \, \underline{\underline{\sigma^+}}\right)_{sym}$$

Where $\underline{\underline{\underline{\underline{i_2}}}}$ is the $\pi/2$ rotation operator in the plan of the material. A scalar effective damage force is build to drive the damage evolution:

$$\overline{z} = max\left(z\left(\underline{\underline{\underline{\underline{Y_C}}}}\right)\right),$$

$$\text{with } z\left(\underline{\underline{\underline{\underline{Y_C}}}}\right) = \left(a\left(tr\left(\widehat{\underline{\underline{Y_C}}}\right)\right)^{n+1} + (1-a)\, tr\left(\widehat{\underline{\underline{Y_C}}}^{n+1}\right)\right)^{1/(n+1)}$$

Where the model parameter $a$ drives the damage anisotropy, the model parameter $n$ being chosen large enough. The associated damage evolution laws are:

$$\begin{cases} \dot{\underline{\underline{\underline{\underline{C}}}}} = \frac{\dot{\alpha}}{\overline{z}^n}\left(a\left(tr\left(\widehat{\underline{\underline{Y_C}}}\right)\right)^n \underline{\underline{1}} + (1-a)\, \widehat{\underline{\underline{Y_C}}}^n\right) \\ \dot{\underline{\underline{\underline{\underline{Z}}}}} = \frac{\dot{\alpha}}{\overline{z}^n}\, b\, \widehat{\underline{\underline{\underline{\underline{Y''}}}}} \end{cases}$$

Where $\alpha(z) = k\sqrt{z}$ is a functional parameter of the model. Let us remark that, $n$ being large enough, $\underline{\underline{\underline{H}}}$ plays the role of an indicator of the damage state –necessary because the damage directions are not known a priori–: it makes the already damaged directions more important than the undamaged ones, so as to drive correctly the damage evolution. This model has been partially identified in [7]. It has already been implemented into the LMT C++ platform [16].

## 3. CHEMICAL PART OF THE MODEL

SiC fibers used in CMCs suffer from sub-critical cracking under oxidizing atmosphere [4]. This is due to defects that propagate from the surface to the core of the fibers (see Fig. 7a). This sub-critical propagation –the normal value of the critical stress intensity factor is not reached on the crack front, but the crack still growths– is driven by the oxidation of the crack front, the arrival of the oxygen being possible by the applied stress that opens the cracks [12]. A new writing of the models of the literature [4, 12], that contains fracture mechanics and is based on a intrinsic law of the reduction of the mechanical properties of the fibers by oxidation, is presented here.

Let us consider a section of the fiber with a surface defect (see Fig. 7b). The size of the defect is written $a$.

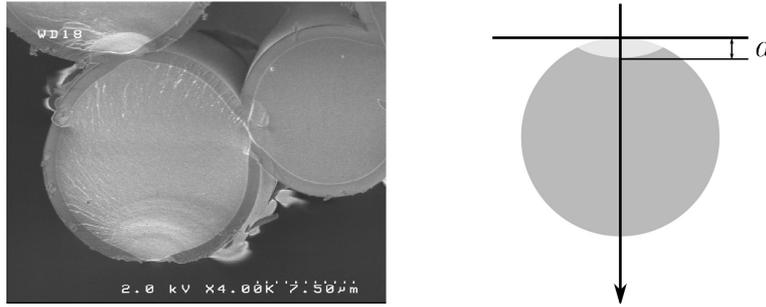

Figure 7: Fiber section with a surface defect, observation and scheme.

This defect induce a stress singularity in the fiber, which stress intensity factor is:
$$K^\sigma(a) = \sigma Y \sqrt{a}$$
Where $\sigma$ is the applied stress on fiber and $Y = 1/\sqrt{2\pi}$ the shape factor [12]. This law gives the value of the initial defect size $a_0$ and the critical defect size $a_c$ in function of the probabilistic strength of the fiber $\sigma_r$ and the applied stress on the fiber $\sigma$:
$$\begin{cases} a_0(\sigma_r) = \left(\dfrac{K_c}{\sigma_r Y}\right)^2 \\ a_c(\sigma) = \left(\dfrac{K_c}{\sigma Y}\right)^2 \end{cases}$$
Where $K_c$ is the undamaged critical stress intensity factor.

In a first step, the propagation of the defect within the fiber section results from the equilibrium between two mechanisms:
- The oxidation of the defect front reduces the local mechanical properties of the fiber, which leads to an increase of its velocity,
- The growth of the defect brings a new crack front, with undamaged mechanical properties, which leads to a slowdown of the defect propagation.

This equilibrium is described by an intrinsic material law that gives the sub-critical stress intensity factor $K_{sc}$ in front of the crack versus the velocity of the crack $v$:

$$K_{sc}^T(v) = \frac{v}{v_0 e^{T/T_0}}$$

Where $v_0$ and $T_0$ are model parameters. The dependence of the temperature $T$ is given in [12]. Inversing this law, one gets the velocity of the defect at a given defect size.

In a second step, one more mechanism has to be introduced:
- The products of oxidation, staying into the defect, limit the arrival of oxygen at the defect front, which leads to a slowdown of the defect propagation.

According to experimental data [12], the defect propagates with a constant velocity during this stage. The transit from first stage to second stage of propagation happens when the defect reach the threshold value $\bar{a}$, which depends on the initial defect size – topology of the defect–, on the applied stress –opening of the defect– , and on the temperature –amount of oxidation products–. As there is no experimental data to clearly identify these influence, the following law is used:

$$\bar{a}^{a_0,T,\sigma} = a_0 + \left(1 - e^{-\left(a_0/a_0^{T,\sigma}\right)^{\alpha_0^{T,\sigma}}}\right)(a_c - a_0)$$

Where $a_0^{T,\sigma}$ and $\alpha_0^{T,\sigma}$ are two model parameters. The last step of propagation starts when the stress intensity factor at the crack front has reached the undamaged critical stress intensity factor. During this stage, the defect propagates almost instantaneously through the whole section, and the fiber is broken. The time to rupture of a fiber is then:

$$t_r(\sigma_r, \sigma, T) = \int_{t(a=a_0(\sigma_r))}^{t(a=a_c(\sigma))} dt$$

$$= \int_{a_0(\sigma_r)}^{\bar{a}^{a_0,T,\sigma}} \frac{da}{v^{T,\sigma}(a)} + \int_{\bar{a}^{a_0,T,\sigma}}^{a_c(\sigma)} \frac{da}{v\left(\bar{a}^{a_0,T,\sigma}\right)}$$

$$= \int_{a_0(\sigma_r)}^{\bar{a}^{a_0,T,\sigma}} \frac{da}{K_{sc}^{T^{-1}}(K^\sigma(a))} + \int_{\bar{a}^{a_0,T,\sigma}}^{a_c(\sigma)} \frac{da}{v\left(\bar{a}^{a_0,T,\sigma}\right)}$$

The predictions of the model are presented Fig. 8, that show the probability of rupture versus time, for different configurations of applied stress and temperature on fibers. Considering the healing of the defects, the model is close to experimental data [4], even for high temperature, which is not the case of the model of the literature [4, 12].

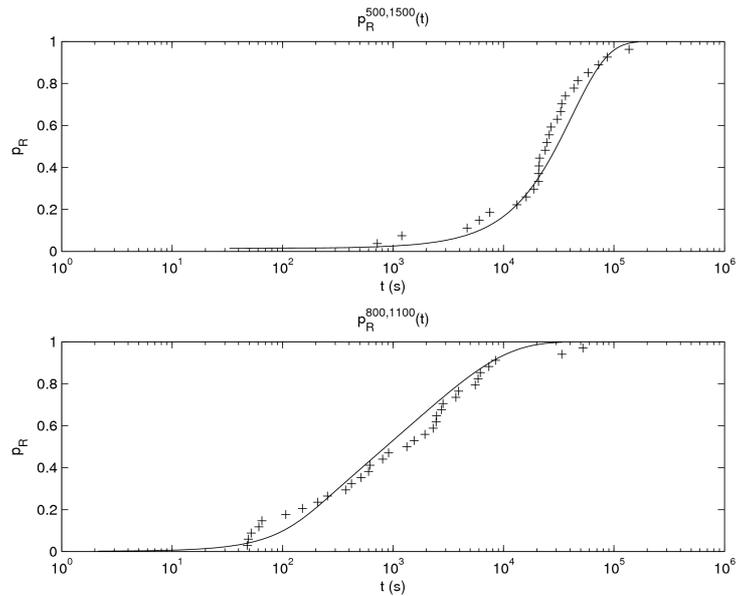

Figure 8: Probability of rupture of fibers under different static fatigue loads, comparison between the model and experimental data [4]

## 4. CONCLUSIONS

A first version of a virtual material for lifetime prediction on self-healing ceramic matrix composites has been presented. The model is divided into two main parts: a mechanical one and a chemical one.

In the mechanical part, the material is principally studied at yarn scale: meshes of meso-cells are created around intra-longitudinal matrix yarns without representing the geometry of the yarns, a dedicated integration procedure linked to level sets functions defined on the mesh allows to take into account complex geometry of woven yarns. The influence of inter-yarn matrix cracking it taken into account through the use of an anisotropic continuum damage model. Intra-yarn matrix cracking appears directly in the calculation through the use of cohesive zone which behavior is given by a micro-macro link. Two important points are to be done now:
- Potential crack surfaces meshing has to be coded in 3D,
- The complete behavior of those potential crack surfaces (fracture criterion, inelastic deformations, crack opening, etc) has to be extracted from the micro-macro link and implemented.

This mechanical part provides the key data for the chemical part: stress on fibers and crack network topology.

In the chemical part, prediction of time to rupture of fibers under mechanical and chemical load are done, based on an intrinsic law of fiber mechanical properties reduction by oxidation. One more process are to be modeled now:
- The self-healing process.

These methodology, once completed, should provide lifetime predictions on self-healing ceramic matrix composites structures.

ACKNOWLEDGEMENTS

The work presented in this paper has been partially funded by French Army through a grant accorded to M. Genet, and by SNECMA Propulsion Solide. The authors would like to specially thank Professor J. Lamon and his students for fruitful collaborations.